\documentclass[twocolumn, switch]{article} 

\usepackage{tablefootnote}

\usepackage{amsmath, amsthm, amssymb, amsfonts}

\usepackage[english]{babel}
\usepackage[autostyle]{csquotes}

\usepackage[numbers,square]{natbib}

\usepackage[utf8]{inputenc}	
\usepackage[T1]{fontenc}	
\usepackage{xcolor}		
\usepackage[colorlinks = true,
            linkcolor = purple,
            urlcolor  = blue,
            citecolor = cyan,
            anchorcolor = black]{hyperref}	
\usepackage{booktabs} 		
\usepackage{nicefrac}		
\usepackage{microtype}		
\usepackage{lineno}		
\usepackage{float}			
\usepackage{graphicx}
\usepackage{lipsum}		

\usepackage{newfloat}
\DeclareFloatingEnvironment[name={Supplementary Figure}]{suppfigure}
\usepackage{sidecap}
\sidecaptionvpos{figure}{c}

\usepackage{titlesec}
\titlespacing\section{0pt}{12pt plus 3pt minus 3pt}{1pt plus 1pt minus 1pt}
\titlespacing\subsection{0pt}{10pt plus 3pt minus 3pt}{1pt plus 1pt minus 1pt}
\titlespacing\subsubsection{0pt}{8pt plus 3pt minus 3pt}{1pt plus 1pt minus 1pt}

\usepackage{listings}
\usepackage{color}

\definecolor{dkgreen}{rgb}{0,0.6,0}
\definecolor{gray}{rgb}{0.5,0.5,0.5}
\definecolor{mauve}{rgb}{0.58,0,0.82}

\usepackage{color}
\definecolor{lightgray}{rgb}{.9,.9,.9}
\definecolor{darkgray}{rgb}{.4,.4,.4}
\definecolor{purple}{rgb}{0.65, 0.12, 0.82}

\lstdefinelanguage{JavaScript}{
  keywords={break, case, catch, continue, debugger, default, delete, do, else, false, finally, for, function, if, in, instanceof, new, null, return, switch, this, throw, true, try, typeof, var, void, while, with},
  morecomment=[l]{//},
  morecomment=[s]{/*}{*/},
  morestring=[b]',
  morestring=[b]",
  ndkeywords={class, export, boolean, throw, implements, import, this},
  keywordstyle=\color{blue}\bfseries,
  ndkeywordstyle=\color{darkgray}\bfseries,
  identifierstyle=\color{black},
  commentstyle=\color{purple}\ttfamily,
  stringstyle=\color{red}\ttfamily,
  sensitive=true
}

\title{AirChain: A Novel Blockchain Framework and Low-Cost Device for Democratized Air Quality Data Aggregation}

\usepackage{authblk}

\author[1]{Samuel Stankiewicz}
\affil[1]{Thomas Jefferson High School for Science and Technology}

\begin{document}

\twocolumn[ 
  \begin{@twocolumnfalse} 
  
\maketitle

\begin{abstract}
Air pollutant exposure kills over 6,700,000 people\cite{fuller2022pollution} it per annum, yet there remains a systemic lack of accurate ground level data reporting the concentrations of the leading causes of such fatalities.  Ambient particulate matter is a primary driver of this effect.  Namely, \texorpdfstring{PM\textsubscript{1.0}, } \texorpdfstring{PM\textsubscript{2.5},} and \texorpdfstring{PM\textsubscript{10.0} d} display a systemic lack of accurate and high-definition reporting. This project suggests and implements a prototype for a distributed and low cost model for reporting such data and designs a novel framework in order to remedy three main shortfalls of previously implemented systems.  First, their central operation and distribution, and therefore their requirement of trust in a central governing body. Second, their requirement of the purchase of comparatively high-cost devices for ordinary consumers.  Finally, their high degree of error and accordingly low functional certainty. This project explores the creation of AirChain, a prototype system utilizing blockchain technology that will demonstrate the effectiveness of low-cost sensors when paired with simple microcontroller devices.
\end{abstract}
\vspace{0.35cm}

  \end{@twocolumnfalse} 
] 



\section{Introduction}
The concept of a distributed air quality monitoring system is not entirely novel.  Namely, several scientists and researchers have proposed similarly designed systems to the one implemented in this paper\cite{jsan9040049} \cite{HAN2019728} \cite{blck1}.  Yet, previous applications require underlying cryptocurrency ties, such as Etherium, and are often limited in scope \cite{MA2023120443}.  Privatized and governmental systems such as the United States’ EPA’s \href{https://airnow.gov}{AirNow} and \href{https://www2.purpleair.com/}{PurpleAir} have demonstrated the effectiveness of a distributed air quality data collection system but all require centralized trust for data authentication and validation. 
 Foreign systems are often segregated and kept within institutional or country-wide walls that lack global consensus.  The creation of AirChain, the networked blockchain system established throughout this paper, will alleviate several main concerns of such systems.  First, the validity of any governmentally-reported or institutional data can be checked by the public using common consumer devices with this architecture.  Contrary to several previous proposals for such systems, this eradicates the need for a completely new system exclusive of all others.  Rather, governmental and private data can be included and stored on the chain just as all other data and flagged according to the reporter.  The scale of the model and an increase in citizen reporting could independently verify the accuracy of institutional data and act as a public check on these data points.  This data is commonly subject to institutional abuse and fraud \cite{Turiel2021-gx}.  Next, the low resolution for many such maps is a paramount concern to unbiased and accurate data reporting.  The EPA’s publicly reported map has been demonstrated to include as few as one sensor per thousand square miles\cite{Coursen.2021}, and the EPA includes no data for sensors measuring \texorpdfstring{PM\textsubscript{1.0}, a p} pollutant attributed to an average 1.67 year reduction in lifespan \cite{Zheng2021}, higher than any other group.  Further, ground level air quality data is commonly substituted with satellite-reported data intended to be representative of ground level concentrations.  Satellite reported air quality data has a standard margin of error as high as 85\% \cite{satallites} due to its high susceptibility to other environmental conditions such as atmospheric diffraction and fog.  Even in best case scenarios, the satalittes were observed to have a standard magin of error between 43-60\%.  The negative effects of air quality exposure are highly correlated with racial and socioeconomic lines on a global scale \cite{RACEAQI}, significantly exacerbating the problem for the world’s poorest.  The exploration of a simple consumer device with the potential to allow ordinary consumers into the data collection system could increase the resolution of such maps by significantly increasing the independent number of total points reported to the network, and therefore this paper seeks to identify a trustless, secure, and reliable method for accurately collecting and distributing this data to and from the public.  The use of a distributed blockchain framework is linear to this motive, and we employ the Hyperledger Sawtooth framework throughout this project to achieve this goal.  A simple user interface conducive to use by the average user could increase the viability and adoption of such a system on a societal scale.

\section{Methodology}
The use of a low-cost device is necessary for this project due to the unique socioeconomic factors which make purchasing expensive laboratory equipment for the ordinary citizen impractical and unreasonable.  The creation of a high scale blockchain system capable of supporting thousands of data points is therefore a significant technical challenge that is uniquely magnified by the low computational capabilities of these devices, including the one explored throughout this paper.  Hyperledger Sawtooth is employed throughout this paper to support the creation of a network with such low technical capabilities.  Sawtooth is an open source permissioned and permissionless Hyperledger project with support for the pBFT, PoET CFT, and PoET SGX consensus algorithms, and developmental support for Raft consensus.  Hyperledger Sawtooth also supports dynamic consensus, a concept explored later in this paper.  These methods of consensus can be used to mitigate environmental concerns such as those introduced by the employment of Proof of Work algorithms common in the cryptocurrency space.  This paper will demonstrate the creation of a novel web interface for the distribution of the air quality data, a companion application for configuring consumer sensors, a basic script for uploading data to an Arduino board capable of running and interfacing with the air quality sensors, and a configuration application for pushing data to the validation nodes which may approve the data to be appended to the chain.  All nodes are written using NodeJS, and the web interface can be independently compiled for public validation and error mitigation.

\subsection{Communication Architecture}
The network utilizes three primary applications, of which any node (defined as an independent machine) has the option to use as many or as few as they would like.  First, a configuration program that allows users with little technical expertise to configure their devices and read their outputs.  This program can also be used to issue API keys via a visual interface without a need for a user to directly submit REST requests to the network.  The API keys are not issued according to a permission schema but are distributed solely for the purpose of having a method of removing and tracking malicious nodes.  

\begin{figure}[!htb]
  \includegraphics[width=\linewidth]{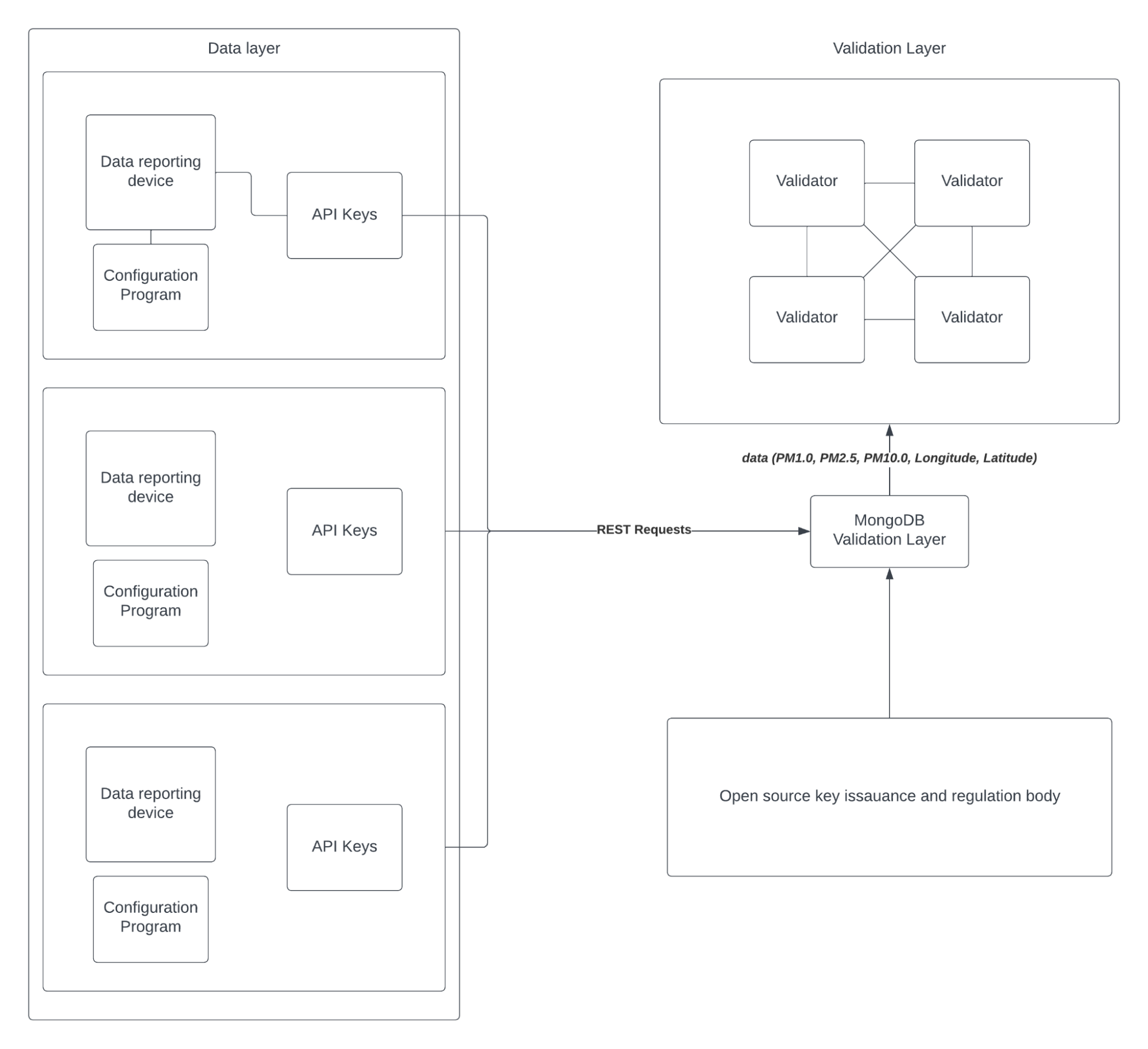}
  \caption{Network architecture}
  \label{fig:basicOutline}
\end{figure}

User accounts are created in a MongoDB database, the contents of which can be made public and managed by an independent consortium of users.  Ideally, the role of such an organization would be limited to revoking API keys and acting as a general check on the network, and clearly this organization would not have the technical ability to modify the contents of any blocks or of the chain itself.  The method by which the chain reaches consensus is explored further in section 2.2.  Validator nodes may be compiled independent of client nodes.  Critically, the computational cost of the system is extremely low.  The sensor employed throughout this experiment, the Plantower PMS7003, has been shown to be capable of operating for 6-8 months before cleaning or maintenance \cite{pmsAccuracy}.  The sensor yields a standard RMSE of 2.22\textmu g/m\textsuperscript{3}, though there may be higher error during months with high heat and low wind.  It has a maximum consistency error of \textpm10\% and can operate between -10{\raise.17ex\hbox{$\scriptstyle\sim$}}60\textdegree C and 0{\raise.17ex\hbox{$\scriptstyle\sim$}}99\% humidity \cite{pms7003}.  Notably, other sensors at slightly higher prices display higher precision \cite{sps30}, but the burden on the user increases and therefore the viability of the system as a whole decreases as the cost of the device increases.  For this reason, alternative sensors were not used for the duration of this experiment.

\subsubsection{Client Configuration Application}
The first of three digital components created for interfacing with the chain is a client application designed for efficiently configuring a low-cost Arduino device connected to a PMS7003 sensor.  The rationale for such a system is relatively simple: the ordinary consumer has no exposure to designing and programming electronics, and the implementation of a system requiring users of the system to connect to and interface with devices themselves would fail.  In short, the implementation of this application is critical for the functioning of the chain and its adoption.  The only prerequisite to downloading and operating it is that the device is connected and has been flashed with the following script:

\begin{lstlisting}
#include "PMS.h"
#include <Wire.h>  
PMS pms(Serial);
PMS::DATA data;

void setup() {
  Serial.begin(9600);
}

void loop() {
  if (pms.read(data)) {
    Serial.print("PM1: ");
    Serial.println(data.PM_AE_UG_1_0);

    Serial.print("PM2.5: ");
    Serial.println(data.PM_AE_UG_2_5);

    Serial.print("PM10: ");
    Serial.println(data.PM_AE_UG_10_0);
  }  
}
\end{lstlisting}

The script could be provided to the public via some common source prior to download.  NodeJS and Electron is employed for the desktop app due to its high flexibility and modularity.    
\\\\The most straightforward way to connect to and interface with the device is the serial interface configured in the installation script.  Following the installation of the desktop app, tethering to the device can be achieved via the following client-side code:

\begin{lstlisting}
Const SerialPort = require('serialport');
port = new SerialPort(req.body.port, {
     baudRate: 9600,
     dataBits: 8,
     parity: 'none',
     stopBits: 1,
     flowControl: false
});

\end{lstlisting}

Then, messages from the device can be intercepted using the following code and sent via webhook:

\begin{lstlisting}
const socket = require('socket.io');
const io = socket(app.listen(3000));
const parsers = SerialPort.parsers;
port.pipe(parser);
const parser = new parsers.Readline({
   delimiter: '\r\n'
});
parser.on('data', function(data) {
     io.emit('data', data);
});
\end{lstlisting}

Alternatively, users can opt to report data from another source or from a preexisting database, both of which would be validated identically to data from PMS sensors.  After the data has been received from the device and interpreted, basic preprocessing using a linear regression model can occur and the data can be pushed to the server via visual interface while utilizing a selected API key from a user account.  Once more, the user account system is necessary to mitigate Distributed Denial of Service attacks and other types of security events and could be managed by an open consortium of users.  Finally, once the level of points has exceeded an arbitrary number and a set amount of time, the application can automatically send a request to the data validators with the matter concentrations and coordinate positions.  The validators can then decide whether to commit the block to the chain or to ignore it.  Critically, requests are sent asynchronously from the client, and the server can respond without needing the client to send a sequential ping.

\begin{figure}[H]
  \includegraphics[width=\linewidth]{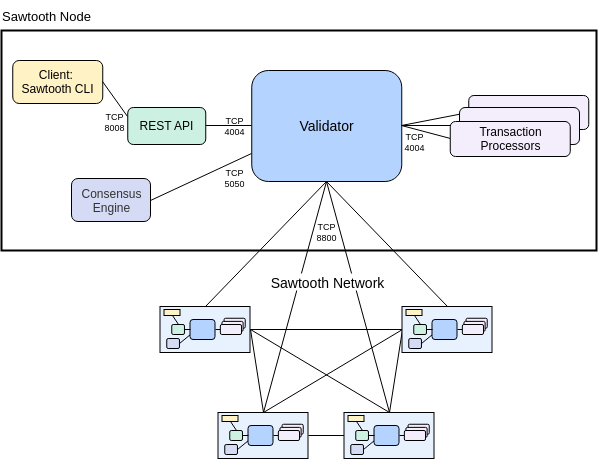}
  \caption{Communications architecture, adapted from \cite{img5}}
  \label{fig:commSchematic}
\end{figure}

\subsubsection{Public Interface Application}

The second component of the proposed framework is a publicly available system for distributing this data.  In theory, this application can also be hosted on a public URL as well as compiled independently such that there is no central dependency on the centrally deployed public website.  The methodology for this site is relatively simple: a server connects to the blockchain and categorically reads all data and plots it on the chart.  A simple timer function can be implemented to ensure that the data remains current, and flags can be implemented to ensure a public user would be aware of the source of any one data point.

\subsubsection{Low-Cost Device}
While many possible methods exist for designing a low-cost device capable of being used by an average user, this paper proposes one built using the previously-mentioned Plantower PMS7003 sensor and an Arduino board. 
 Flashing the device is relatively straightforward, and little electronics experience is required to configure the controller.  Notably, the device could also be attached to an ESP32 board and could communicate without the need for a tether to an external device if the user connected the board to an external power supply such as a battery pack.  Of course, another method for device communication would be needed in this situation as the serial interface would no longer function.

\begin{figure}[H]
  \includegraphics[width=\linewidth]{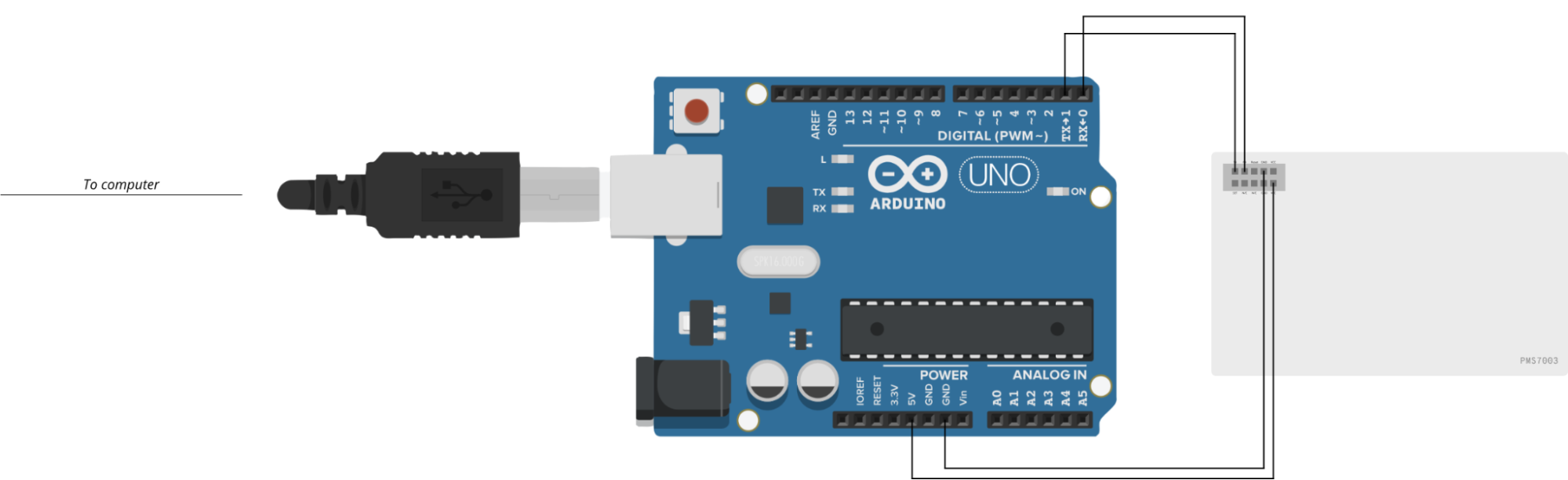}
  \caption{Basic sensor schematic}
  \label{fig:arduinoSchematic}
\end{figure}

\subsection{Chain Consensus}
As previously explored, Sawtooth is capable of dynamic consensus, meaning that the consensus algorithm of the chain can be changed at will and is even modifiable after the creation of a genesis block and the operation of the network.  Therefore, this paper is agnostic of the specific algorithm used but proposes several ideas as to how such a network could be operated and will explore the benefits and drawbacks of each algorithm and how they might be used in this system. 

\subsubsection{Practical Byzantine Fault Tolerance (pFBT)}
Practical Byzantine Fault Tolerance, or pBFT, is a non-forking algorithm that may be the most straightforward method of validating the data on the chain due to several factors but ultimately likely lacks the necessary scale for a true public deployment.  First, pBFT only requires reliance on 3m + 1 trustworthy nodes \cite{complexity} where m is the number of malicious nodes.  This is a major advantage for a public system in which the validity of any one actor cannot be independently verified.  However, pBFT involves a high level of communication between nodes and validators and is therefore extremely computationally expensive at scale.  Creating a high scale pBFT network would likely be extremely difficult, and therefore the use of a pBFT consensus algorithm for this project is not recommended.

\subsubsection{Proof of Elapsed Time (Crash Fault Tolerant)}
Proof of Elapsed Time with Crash Fault Tolerance (PoET CFT) is proprietary Nakamoto-style consensus algorithm unique to the Sawtooth platform that implements a lottery-style validation system.  PoET CFT is highly fair and implements a random lottery-style leader selection method.  PoET CFT is non-forking and is primarily permissioned.  This system has several advantages when compared to a pBFT system.  First, PoET CFT can operate in the presence of faulty or technically hampered nodes.  This is a significant advantage to a system reliant on blocks published by consumer devices.  However, PoET CFT is not Byzantine Fault Tolerant, and this therefore makes the use of the algorithm difficult to justify.  Many experts have raised concerns about PoET’s vulnerability to Sybil attacks, which may be a critical concern for a system with a theoretically high public impact.  The Sawtooth platform preforms Z tests to determine if a particular node is winning at a rate significantly higher than the mean win rate, yet these tests often fall 

\subsubsection{Proof of Elapsed Time (Byzantine Fault Tolerant on Intel SGX)}
The use of a Proof of Elapsed Time algorithm with Byzantine Fault Tolerance (PoET SGX) is another significant possibility for a public network such as the one proposed in this paper, and in this sense it may be a more effective consensus algorithm for this project when compared to pBFT or PoET CFT.  PoET SGX is non-forking and is also primarily permissioned like PoET CFT.  PoET SGX has similar vulnerability to Sybil attacks, but also has a unique vulnerability to physical cracks due to its reliance on Intel SGX processors.  Various experts have explored and exploited these faults \cite{poetVulnerability} \cite{Wang2022MultiCertificateAA}.  PoET SGX has very low computational requirements but requires the use of an Intel processor, and this may not be conducive to a high-scale public model.  One major disadvantage of reliance on an Intel infrastructure is that this effectively decentralizes the network and creates a central reliance on the Intel Corporation.  PoET SGX also has relatively high transaction times, which may not be conducive to a high scale network.  Finally, PoET SGX is theorically vulnerable to a crack of as little as
\[\Theta(\dfrac{(\log \log n)}{\log n})\]
nodes, an extremely small number at scale.  This method of attack could prove deadly to a network of this type.  Sawtooth and REM have argued that this method of attack can be prevented via Z testing of all nodes, a claim disputed in some network hacks \cite{Wang2022MultiCertificateAA}.

\subsubsection{Raft Consensus}
Raft is another promising consensus algorithm that implements an elected-leader based authentication model.  A primary advantage of Raft is that it is significantly faster than PoET, but this algorithm is Crash Fault Tolerant and not Byzantine Fault Tolerant, a concern that may be critical to a high scale application.  All place authoritative trust in the leader node in a Raft consensus algorithm.  While Crash Fault Tolerance may be an important component of a functional consensus mechanism, Raft’s lack of Byzantine Fault Tolerance may make the practical feasibility of such an algorithm low in this application.  Additionally, Raft requires fixed membership and does therefore not support the adding of new nodes without administrator configuration.  Finally, Raft requires a significant amount of messaging between nodes for validation and is therefore not conducive to a large scale network, in which the computational time conforms to an exponential curve \cite{UCAM-CL-TR-857}. Summarily, this method of consensus would be largely infeasible if the network exceeded a total of ten nodes, a relatively low number for the given application.

\section{Network Validation Architecture}
Regardless of the consensus mechanism used, the architecture of the network remains largely consistent between peers.  This section will outline the architecture of each component of the system.

\subsection{Visual Nodes}
\begin{figure}[H]
  \includegraphics[width=\linewidth]{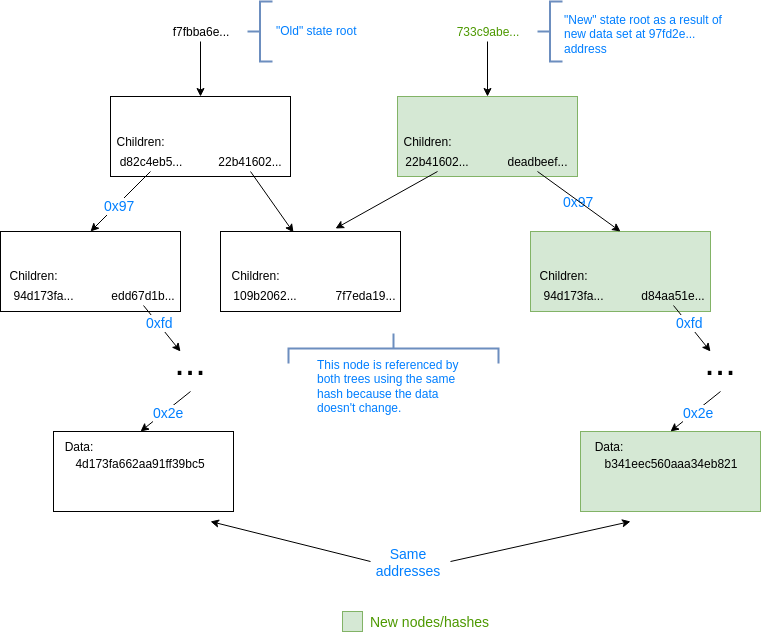}
  \caption{Merkle-radix tree in a Sawtooth application, adapted from \cite{img3}}
  \label{fig:merkleSchematic}
\end{figure}
The visual interface used for displaying the data acts as a validator node in our system such that all incoming requests are parsed via the REST API and validated by the validator component using an arbitrary consensus mechanism.  REST requests from the client are sent to the API over TCP port 8008, and requests are then passed to the validator over TCP port 4004.  The consensus engine communicates with the validator over TCP port 5050.  Finally, the validator interfaces with the transaction processors over TCP 4004 to perform operational tasks and other configuration.  This process is described in figure \ref{fig:merkleSchematic}.  After the validation of a block, it is stored in a merkle-radix tree in which the immutability and security of the chain can be assured.

\subsection{Peering and Network Discovery}

\begin{figure}[H]
  \includegraphics[width=\linewidth]{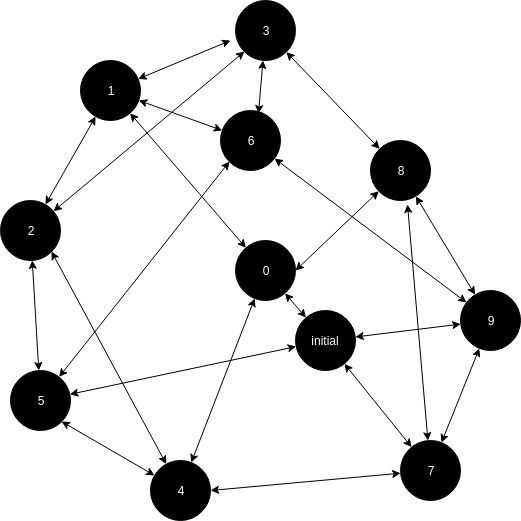}
  \caption{Network peering, adapted from \cite{img8}}
  \label{fig:peeringArchitecture}
\end{figure}

Peering is the process by which block nodes discover each other and connect in situ.  Each node acts as a peer on the network, and the process of peering is performed by sending sequential CONNECT/GET\textunderscore PEERS requests to peer nodes on the network until every node has reached minimum connectivity.  If a node has reached the maximum allowed connectivity, the peering process continues and nodes continue to attempt connection to each other.  This process recurs until every node on the network has reached minimum connectivity or each unpeered neighbor has sent peering requests to every available node on the network.  At this point, a successful network is considered peered. 

\subsection{Transactions, Journals, and Validator Journals}
The client REST API submits Sawtooth transactions, which become wrapped into batches.  Batches are serialized into batch headers, and transactions are serialized into transaction headers.  Both components are signed using a secp256k1 elliptical curve.  Payloads are validated using SHA-512 hashes.  The validator validates batches and pushes them to the head of the blockchain.  This process is contained within a validator journal, groups of validator subcomponents.  Collectively, they manage the flow of proposed blocks, and they push blocks to the chain while validating proposed blocks.  
\begin{figure}[H]
  \includegraphics[width=\linewidth]{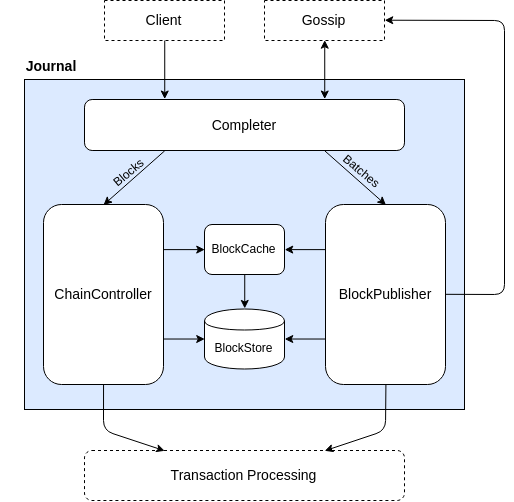}
  \caption{Journal structure, adapted from \cite{img7}}
  \label{fig:journalStructure}
\end{figure}
All blocks must undergo validation and pass through the validation journal process to be eligible to add to the chain.  This section describes validation journals and explains how they are used in the Sawtooth network to approve and validate new blocks as well as add them to the chain head.

\begin{itemize}
  \item The completer element receives incoming blocks from the client or via the gossip channel and ensures that all needed dependencies for the block to undergo publishing are met.  After this validation has occurred, approved blocks are passed to the Chain Controller, and approved batches are passed to the Block Publisher.
  \item The Chain Controller element validates incoming blocks from the completer and resolves forks in the chain.  It sends approved blocks to the Block Cache, the Block Store, and to transaction processing.  The text in the entries may be of any length.
  \item The Block Publisher receives batches from the completer and validates them for inclusion in new blocks.  It sends these to the Block Cache, the Block Store, the Gossip channel, and to transaction processing after completion 
  \item Both the Block Cache and the Block Store store batches and blocks during validation
\end{itemize}

\section{Discussion and Conclusion}
The establishment of a proof-of-concept blockchain network capable of storing public air quality data at scale was successful during this project, and the use of a Hyperledger Sawtooth framework was successfully explored.  Further research is needed into possible data validation methods for this distributed method of air quality authentication, including especially the detection and removal of fraudulent and malicious nodes within the network.  In theory, the implementation of a Convolutional Neural Network or other Artificial Intelligence model class could eliminate the need for an open source consortium of human managers, and this would allow for another correlative reduction in the centralization of the network.  Finally, further exploration and research into how such a system could be adopted on a societal scale is needed, and methods for exploring how to effectively communicate a network of this type to the public are a necessity for continued action in this field.

\section{Acknowledgements}
This research was conducted exclusively by Samuel Stankiewicz, a student at Thomas Jefferson High School for Science and Technology.  No grants or other funding was received for this project.

\normalsize
\bibliographystyle{plain}
\bibliography{references}
\end{document}